# Co-ordering and Type 2 co-ordering

**Saeed Asaeedi**
(AmirKabir University of Technology, Tehran, Iran
sa_saeed_sa@aut.ac.ir)

**Farzad Didehvar**
(AmirKabir University of Technology, Tehran, Iran
didehvar@aut.ac.ir)

**Introduction:**

In [1], [2] the enumeration order reducibility is defined on natural numbers. For a c.e. set $A$, $[A]$ denoted the class of all subsets of natural numbers which are co-order with $A$. In definition 5 we redefine co-ordering for rational numbers.

One of the main questions there, was:

*"For a specific c.e. set A, consider set of all enumerations of it which is generated by some Turing machine $TM_A$ what are the associated order types in [A]?"*

Here, we propose the same question for rational numbers, and we try to investigate the varieties of c.e. sets on $\mathbb{Q}$. The theories here are hold for $R_c$ and we could repeat the same theories in this domain, in a parallel way.

First, we define the basic concepts again, in the second steps we are going to prove Theorems and simultaneously we compare these Theorems to the Theorems in the subject of Enumeration order reducibility on natural numbers.

**Definition 1:** The Function $g: \subseteq \mathbb{N} \to \mathbb{Q}$ is computable if there exist computable functions $a, b, s: \subseteq \mathbb{N} \to \mathbb{N}$ such that $\forall i \ b(i) \neq 0$ and $g(n) = (-1)^{s(n)} \frac{a(n)}{b(n)}$.

**Definition 2:** The set $A \subseteq \mathbb{Q}$ is recursive if the characteristic function of $A$ ($\chi_A$) be computable

$$\chi_A(x) = \begin{cases} 1 & x \in A \\ 0 & o.w \end{cases}$$

**Definition 3:** The set $A$ is c.e. if there exists a computable function like $f$ such that $A = Range(f)$.

**Example:**

1. $A = \{x \in \mathbb{Q} | \ a \leq x \leq b\}$ is recursive
2. $B = \{2^{-m} | \ m \in A\}$ is non-recursive c.e. set such that $A \subseteq \mathbb{N}$ is non-recursive c.e. set.

**Definition 4:** A listing of an infinite c.e. set $A \subseteq \mathbb{Q}$ is a bijective computable function $f: \mathbb{N} \to A$.

**Definition 5:**

- Two listings $h$, $g$ are **co-order**, $h \sim g$, if $h(i) < h(j) \Leftrightarrow g(i) < g(j)$ for all $i, j \in \mathbb{N}$.
- Two c.e subsets of $\mathbb{Q}$, $A$ and $B$, with equal cardinality are **co-order** ($A \sim B$) if there exist listings $h$ of $A$ and $g$ of $B$ such that $h \sim g$.

For a c.e. set $A \subseteq \mathbb{Q}$, let $[A]_\sim$ denotes the Enumeration order equivalence class of $A$, ($[A]_\sim = \{B \subseteq \mathbb{Q} | A \sim B\}$).

**Example:**

1. Any two finite recursive set $C, D \subseteq \mathbb{Q}$ with the same cardinality are **co-order**.
2. $A = \{\frac{1}{n} | \ n \in \mathbb{N}\}$ and $B = \{\frac{n}{3} | \ n \in \mathbb{N}\}$ are not **co-order.**
3. K and $\mathbb{N}$ are not **co-order**.

**Theorem 1:** The following statements are equivalent:

1. $A \sim B$.
2. For every listing $h$ of $A$ there is a listing $g$ of $B$ such that $h \sim g$.

**Theorem 2:** There are two recursive sets $A$ and $B$ subsets of $\mathbb{Q}$, such that they are not co-order.(In contrast to this fact that any two recursive set which are subsets of natural numbers are co order).

## Co-ordering and One-reducibility

By following lemma we consider the relation between Enumeration order equivalency classes and One-reducibility:

**Lemma 1:** let two sets $A, B \subseteq \mathbb{Q}$ be c.e. sets. $A$ and $B$ do not belong to the same Enumeration order equivalence class necessary, if they belong to the same one-reducibility class.

**Proof:** Just consider two sets $A = \{\frac{1}{n} | \ n \in \mathbb{N}\}$ and $B = \{\frac{n}{3} | \ n \in \mathbb{N}\}$.

**Lemma 2:** let two sets $A, B \subseteq \mathbb{Q}$ be c.e. sets. $A$ and $B$ do not belong to the same one-reducibility class necessary, if they belong to the same Enumeration order equivalence class.

### Co-ordering and Turing-reducibility

In the following lemma we investigate the relation between Enumeration order equivalency classes and Turing-reducibility:

**Lemma 3:** let two sets $A, B \subseteq \mathbb{Q}$ be c.e. sets. $A$ and $B$ do not belong to the same Enumeration order equivalence class necessary, if they belong to the same Turing-reducibility class.

**Proof:** By lemma 1, the lemma 3 is immediate.

**Lemma 4:** let two sets $A, B \subseteq \mathbb{Q}$ be c.e. sets. If $A, B$ belong to the same Enumeration order equivalence class, they belong to the same Turing-reducibility equivalence class.

**Theorem 3:** let $A$ be a c.e. set. There are infinitely many c.e. sets $B_i$ such that $[B_i]_\sim \subseteq [A]_T$.

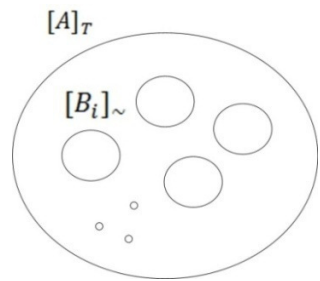

### Type 2 co-ordering

Adding or removing finite subset of a non-recursive c.e set like $A$, leads us to create a c.e set that is not co-order with $A$. So, we define the type 2 co-ordering here.

**Definition 6:** Two sets $A$ and $B$ are almost equivalent ($A \approx B$) if $A \nabla B$ be finite.

**Definition 7:** Two sets $A$ and $B$ are type2 co-order ($A \sim_2 B$) if there exist two sets like $C, D$ such that $A \approx C$ and $B \approx D$ and $C \sim D$.

**Lemma 4:** let two sets $A, B \subseteq \mathbb{Q}$ be c.e. sets. $A \sim_2 B$ iff there exist finite sets $C, D \subseteq \mathbb{Q}$ such that $A - C \sim B - D$.

**Definition 8:** Two listing $h, g$ are type2 co-order ($h \sim_2 g$) if there exist $m, n \in \mathbb{N}$ such that for all $\in \mathbb{N}$, $h'(i) = h(i + m)$ and $g'(i) = g(i + n)$ and $g' \sim h'$.

**Theorem 4:** For infinite c.e. sets $A, B$ the following statements are equivalent:
   1. $A \sim_2 B$.

2. There exist listings $h$ of $A$ and $g$ of $B$ such that $h \sim_2 g$.
3. For any listing $h$ of $A$ there exists a listing $g$ of $B$ such that $h \sim_2 g$.

Like co-ordering relation, type2 co-ordering is an equivalence relation. For a c.e. set $A \subseteq \mathbb{Q}$, let $[A]_{\sim_2}$ denotes the Type2 enumeration order equivalence class of $A$, ($\{B \subseteq \mathbb{Q} | A \sim_2 B\}$).

**Notation:** In the subsequent lemmas and theorems, we utilize the following notations for every two listings $h$ and $g$:
1. For all $m, n \in \mathbb{N}$ the notation $E_{m,n}^{h,g}$ denotes the set
$$E_{m,n}^{h,g} = \{(i,j) \in \mathbb{N}^2 | h(i+m) < h(j+m) \text{ and } g(i+n) > g(j+n)\}$$
2. For all $m, n \in \mathbb{N}$ the notation $M_{m,n}^{h,g}$ denotes the set
$$M_{m,n}^{h,g} = \{i \in \mathbb{N} | there\ exist\ a\ j \in \mathbb{N}\ such\ that\ (i,j) \in E_{m,n}^{h,g}\}$$
3. For all $m, n \in \mathbb{N}$ the notation $L_{m,n}^{h,g}$ denotes the set
$$L_{m,n}^{h,g} = \{j \in \mathbb{N} | there\ exist\ a\ i \in \mathbb{N}\ such\ that\ (i,j) \in E_{m,n}^{h,g}\}$$

**Lemma 5:** If listings $h, g$ are not type2 co-order, then for all $m, n \in \mathbb{N}$, $M_{m,n}^{h,g}$ and $L_{m,n}^{h,g}$ are infinite.

**Theorem 5:** Let c.e. sets $A, B$ are not type2 co-order. Consider an c.e. set $C$ such that $A \nsubseteq C$. $A \cup C$ and $B$ are not type2 co-order.

**Type 2 co-ordering and Reducibility**

Here we survey the relation between type2 co-ordering and one-reducibility and Turing-reducibility.

**Theorem 6:** If two sets belong to the same one-reducibility equivalence class, then they do not necessarily belong to the same Type2 enumeration order equivalence class.

**Theorem 7:** Two type2 co-order sets do not belong necessarily to the same one-reducibility equivalence class.

**Theorem 8:** If two c.e. sets $A$ and $B$ are type2 co-order then they belong to the same Turing-reducibility equivalence class.

In natural numbers, by considering $[\emptyset]_T$, the equivalence class of recursive sets, it was proved in [1] that There are only two type2 enumeration order equivalence classes such that they are subsets of the equivalence class $[\emptyset]_T$. Let we consider the following theorem in rational numbers:

**Theorem 9:** There are infinite numbers of Type2 enumeration order equivalence classes such that they are subsets of equivalence class $[\emptyset]_T$.

**Proof:** Let we define $T_i$ for all $i \geq 1$:
$$T_i = \begin{cases} (i-1) + \frac{n-1}{n} & if\ i\ is\ odd \\ i - \frac{n-1}{n} & if\ i\ is\ even \end{cases}$$

If $i$ is odd, $T_i$ is ascending And if $i$ is even, $T_i$ is descending. The following figure shows the behavior of $T_i$:

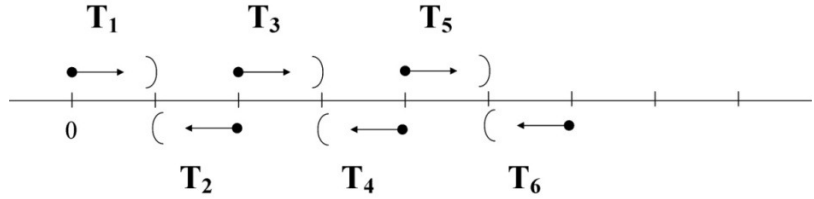

We define the recursive sets $A_i$ as $A_i = \bigcup_{s=1}^{i} T_s$.
We must show that for all $i, j \in \mathbb{N}$ that $i \neq j$ we have $A_i \not\sim_2 A_j$.
It's simple to show that for all $i \in \mathbb{N}$, $A_i \not\sim_2 T_{i+1}$.
First we show for all $i \in \mathbb{N}$ that $i \neq 1$, $A_1 \not\sim_2 A_i$. We use theorem 5 here:
- $A_1 = T_1 \not\sim_2 A_2$
- $A_1 \not\sim_2 A_2$ , $A_2 \nsubseteq T_3$ $\Rightarrow$ $A_2 \cup T_3 \not\sim_2 A_1$ $\Rightarrow$ $A_3 \not\sim_2 A_1$
- $A_1 \not\sim_2 A_3$ , $A_3 \nsubseteq T_4$ $\Rightarrow$ $A_3 \cup T_4 \not\sim_2 A_1$ $\Rightarrow$ $A_4 \not\sim_2 A_1$
- .
- .
- $A_1 \not\sim_2 A_i$ , $A_i \nsubseteq T_{i+1}$ $\Rightarrow$ $A_i \cup T_{i+1} \not\sim_2 A_1$ $\Rightarrow$ $A_{i+1} \not\sim_2 A_1$

For $A_2$ we just need to show $A_2 \not\sim_2 A_3$. In the same way we can show for all $i \in \mathbb{N}$ that $i \neq 2$, $A_2 \not\sim_2 A_i$.
$A_2 \not\sim_2 T_3$ $\Rightarrow$ $A_2 \not\sim_2 T_3 \cup A_2$ $\Rightarrow$ $A_2 \not\sim_2 A_3$.
Analogously, for $A_3, A_4, \ldots$ ,we can show for all $i, j \in \mathbb{N}$ , $A_i \not\sim_2 A_j$. ∎

**Theorem 10:** Let $A$ be a non-recursive c.e. Set. There are infinite numbers of Type2 enumeration order equivalence classes such that they are subsets of equivalence class $[A]_T$.

**Proof:** Since recursively enumerable Turing degrees are dense, there is a chain of non-decidable c.e. sets $\{A_i\}_{i \in \mathbb{N}}$ such that for all $i \in \mathbb{N}$, $A_{i+1} <_T A_i$ and $A_0 = A$ and for distinct members of this sequence $A_r$ and $A_s$, $A_r \nsubseteq A_s$. For all $i, j \in \mathbb{N}$, $[A_i]_T \neq [A_j]_T$ and thus based on the theorem 8 $[A_i]_{\sim_2} \neq [A_j]_{\sim_2}$. Let $B = <B_i>_{i \in N}$ be an increasing chain such that $B_i = \bigcup_{0 \leq k \leq i} A_k$. For all $i \in \mathbb{N}$, $B_i \in [A]_T$. According to the theorem 5 for all $i, j \in \mathbb{N}$ and $i \neq j$, $B_i$ and $B_j$ are not type2 co-order. So there are infinite numbers of Type2 enumeration order equivalence classes such that they are subsets of the equivalence class $[A]_T$. ∎

**Acknowledgment:**
We would like to thank the staff of department of mathematics and computer science and thank Aliakbar Safilian for some of his cooperations to prepare this paper.